# Covid-19 Tweeting in English: Gender Differences

Mike Thelwall, School of Mathematics and Computer Science, University of Wolverhampton, UK.
Email: m.thelwall@wlv.ac.uk. Orcid: 0000-0001-6065-205X
Saheeda Thelwall, Institute of Health, University of Wolverhampton, UK.
Email: s.thelwall@wlv.ac.uk Orcid: 0000-0002-0333-399X

At the start of 2020, COVID-19 became the most urgent threat to global public health. Uniquely in recent times, governments have imposed partly voluntary, partly compulsory restrictions on the population to slow the spread of the virus. In this context, public attitudes and behaviors are vitally important for reducing the death rate. Analyzing tweets about the disease may therefore give insights into public reactions that may help guide public information campaigns. This article analyses 3,038,026 English tweets about COVID-19 from March 10 to 23, 2020. It focuses on one relevant aspect of public reaction: gender differences. The results show that females are more likely to tweet about the virus in the context of family, social distancing and healthcare whereas males are more likely to tweet about sports cancellations, the global spread of the virus and political reactions. Thus, women seem to be taking a disproportionate share of the responsibility for directly keeping the population safe. The detailed results may be useful to inform public information announcements and to help understand the spread of the virus. For example, failure to impose a sporting bans whilst encouraging social distancing may send mixed messages to males.

## Introduction

COVID-19 is, at the time of writing, a major global threat to public health (e.g., Lipsitch, Swerdlow, & Finelli, 2020). Public actions are critically important in slowing the spread of the virus and therefore reducing the death rate due to the volume of critically ill patients needing simultaneous care, such as by running out of ventilators. Governments around the world have reacted by announcing mandatory actions, such as shutting restaurants and the normal functioning of schools, and by giving strongly recommended or mandatory advice to the public for personal hygiene and social distancing to slow the spread of the virus. The extent to which the population follows expert health advice is expected to have a substantial impact on the death rate from the virus. If social distancing is widely ignored or misunderstood, for example, then national healthcare facilities will not be able to give all critically ill patients the care that they need to survive. It is therefore vitally important to assess how the public is reacting to the crisis and one way (amongst many) of investigating this is through social media posts, including tweets (e.g., Cinelli, Quattrociocchi, Galeazzi, et al., 2020), and one important potential arena of difference (amongst many) is gender.

Twitter is a natural platform for public information sharing in many countries, including all large English-speaking nations. Although less popular than Facebook, its advantage for research is that it is typically fully public and researchers can therefore access its contents. Moreover, Twitter gives free use of an Applications Programming Interface (API) for automatically harvesting recent (up to a week old) tweets matching keyword searches, making it a practical source of data about public reactions to tweets. A disadvantage is that Twitter users' demographics do not match those of the population. In the USA, about 23% of adults use the site, behind Facebook (71%) and Instagram (38%), but ahead of WhatsApp

(18%) and Reddit (13%) (Schaeffer, 2019). Moreover, older people (and more at risk from COVID-19) are be less likely to use Twitter, men are slightly more likely to use it (50% female within a 52% female population) but adopters tend to be richer and more educated in the USA (Smith & Wojcik, 2019). There are also finer-grained differences, such as political variations between users and non-users (Smith, Hughes, Remy, & Shah, 2020). Nevertheless, analyzing tweets may give some quick large-scale insights into public reactions to COVID-19.

This study focuses on gender differences in reactions to COVID-19 on Twitter. Since public safety measures must be adhered to by the entire population to be maximally effective, any gender differences in responses may point to weaknesses in public communications about the seriousness of the outbreak. This information may help with the creation of new messages targeting males or females more effectively. In addition, understanding gender differences may help modelling epidemiologists to create more accurate models of the spread of the disease. The current paper therefore analyses two weeks (March 10-23, 2020) of tweeting in English about COVID-19 from the perspective of gender differences in responses. Although the virus is a global pandemic, the focus on English is for pragmatic methodology reasons and similar research in other languages is encouraged (and supported by the free software at http://mozdeh.wlv.ac.uk).

## Methods

The research design was to collect English-language tweets matching a set of queries related to Covid-19 over two weeks and to identify words used more by males than females, using these to point to aspects of gender difference in tweeting about the virus. A word frequency method is useful for gender comparisons because it gives statistically significant evidence in a transparent fashion. In contrast, content analysis or thematic analysis are unlikely to discover fine-grained gender differences and cluster-based methods, such as topic modelling, can be changed by small alterations in the data, and so are not robust. Topic modelling is also not able to give as fine-grained gender difference information as word frequency comparisons. Word frequency analysis therefore fills a gap in comparison to other methods.

The following queries were used to identify different common ways of referring to the disease: coronavirus; "corona virus"; COVID-19; COVID19. These were submitted to Twitter at the maximum speed allowed by the free Twitter API from 10 to 23 March 2020, obtaining 3,038,026 tweets after eliminating duplicates (including multiple retweets) and near duplicates (tweets identical apart from @usernames and #hashtags). The tweets were collected and analyzed with the free software Mozdeh (http://mozdeh.wlv.ac.uk).

Twitter does not record user genders, but it is possible to guess male and female genders (only) from their display name if it starts with a first name. A list of gendered first names was used to match the first part of Twitter display names. This list was US-based, since the USA is the major English-language user of Twitter and its population has international ethnic origins, so its names probably reflect to some extent the names in other anglophone countries. The list was derived from the 1990 US census (top 10000 names) and supplemented by GenderApi.com (names with at least 100 US records). Names were included as female (respectively, male) from either source if at least 90% of people with the name were female (respectively, male). Twitter names (display names, rather than usernames) were split at the first space or non-alphanumeric character, first digit, or first camel case transition from lowercase to uppercase (e.g., MikeThelwall). The 90% threshold was chosen to give a high degree of certainty that the user was male. The method is imperfect because Twitter usernames may be informal or not reflect a person's name (e.g., CricketFan938624), or based on a relatively gender-neutral name (e.g., Sam, Pat) or a rare name, including names from small ethnic minorities in the USA. Nevertheless, the first name procedure splits a set of tweets into three groups: (a) likely to be male-authored; (b) likely to be female-authored; (c) unknown. Comparing (a) with (b) gives an indication of likely gender differences overall. Visual inspection of the most active users in the data suggests that most bot and corporate tweets are assigned to the unknown gender set.

Gender differences in topics were identified by a word frequency comparison method to identify words more used by either males or females, using the following procedure. For each word, the proportion of female-authored tweets containing the word was compared to the proportion of male-authored tweets containing the word using a 2x2 chi-square test for the table: [[Female tweets with word, Female tweets without word],[Male tweets with word, Male tweets without word]]. A statistically significant chi-squared value (3.841 for p=0.05) gives evidence to reject the null hypothesis of no gender difference in use of the word. Because the test is repeated for every word and there are 1,372,497 words, this procedure would almost certainly produce tens of thousands of false positives due to the number of tests. The Benjamini-Hochberg procedure (Benjamini & Hochberg, 1995) was used to correct for this. It is a familywise error rate correction procedure that ensures that the probability of incorrectly rejecting the null hypothesis in any test is below a threshold value. For extra power, words that were too rare to trigger a statistically significant result, even they were only used by males (or females) were not tested. This chisquared/Benjamini-Hochberg approach for detecting gender differences in term frequencies has previously been used for academic abstracts (Thelwall, Bailey, Makita, Sud, & Madalli, 2019; Thelwall, Bailey, Tobin, & Bradshaw, 2019), Reddit posts (Thelwall & Stuart, 2019) and YouTube comments (Thelwall, 2018). The procedure was repeated three times, for p=0.05, p=0.01, and p=0.001, recording the highest significance level for each word.

The above procedure was also applied to each day separately to determine the statistically significantly gendered terms for each day (i.e., 14 additional sets of tests). This extra step was taken because a word that is gendered on a single day seems likely to be less relevant to Covid-19 than a word that is gendered on multiple days. For example, a one-day gendered term might relate to a news event that was affected by Covid-19 (e.g., a sporting event cancellation) but this might not be important to the ongoing discussion of the virus. The threshold for including a term was set at (the equivalent of) more than two highly statistically significant days. Allocating one star to significance at p=0.05, two for p=0.01 and three for p=0.001, the threshold requirement was a total of at least seven stars over the fourteen days. This threshold gave a total of 102 terms out of the 339 that were statistically significantly gendered on at least one day.

Each word judged statistically significantly gendered (either overall, or on multiple days) reflects one or more underlying gender differences in motivations for tweeting or a gender difference in language styles. Each term's underlying causes can be inferred by reading a random sample of tweets containing the term, known as the Key Word In Context (KWIC) method (Luhn, 1960). For example, the term *league* was associated with tweets discussing the full or partial closure of various sporting competitions or facilities. Gender differences in this word therefore suggest that males were more likely to tweet that league-based sport was affected by COVID-19 restrictions. The word contexts varied from obvious (e.g., #jantacurfew) to obscure (e.g., it). In particular, many pronouns were female associated, reflecting a people-focus rather than a topic, and definite and indefinite articles were male-associated, reflecting an information focus rather than a specific topic. In cases where the context of a term was unclear from reading ten randomly selected tweets (using the random sort option in Mozdeh), a word association analysis was run on the term to identify top associating terms to give additional insights into its main use contexts.

The words were manually grouped into themes for each gender to highlight the main types of gender difference.

## Results

The main themes identified in the tweets are summarized below by gender. The complete list of terms and raw tweet counts associated with them are available on FigShare (https:doi.org/10.6084/m9.figshare.12026625).

### Male-oriented themes

Male-authored tweets about COVID-19 were about twice as likely as females to discuss sports, typically in the context of speculation about, or announcements of, events or competitions being cancelled (Figure 1). Whilst this is relatively peripheral to the disease, males were also substantially more likely to mention, or take issue with, political figures or government, particularly within India (Figure 2). Males were also more likely to tweet about the economy (terms: economy, market; not graphed).

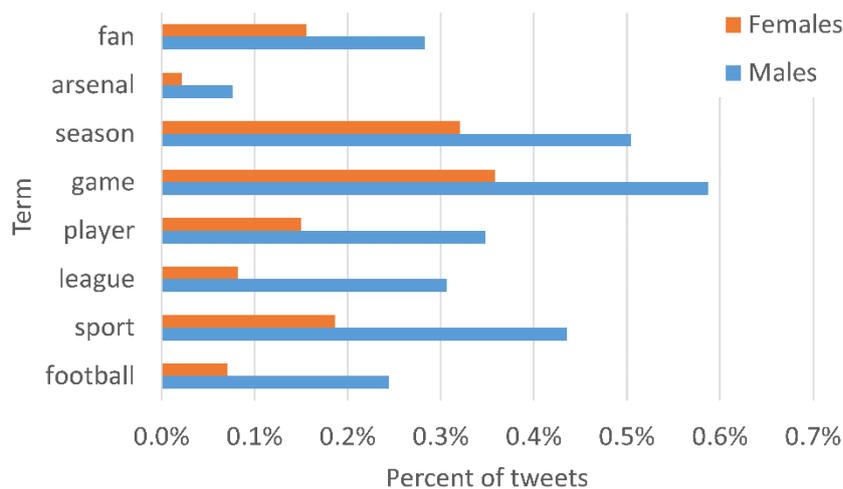

Figure 1. Sport-related terms with gender differences in usage.

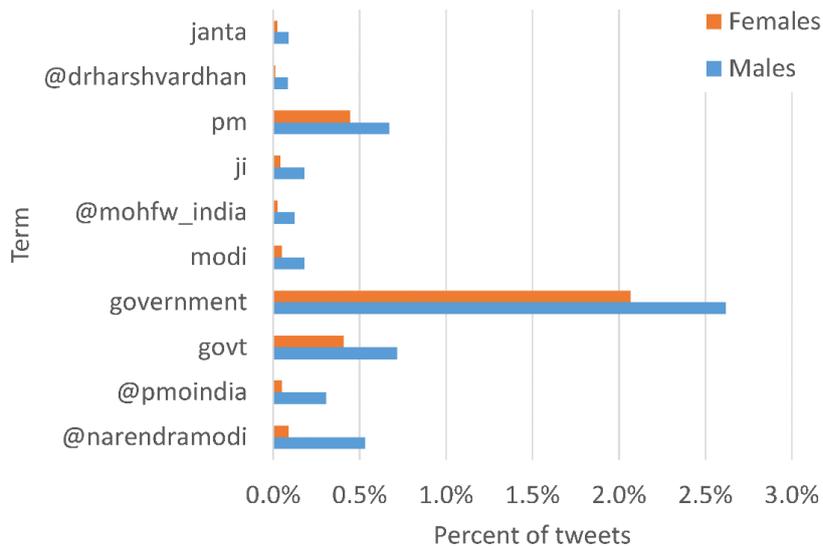

Figure 2. Politics-related terms with gender differences in usage.

The epidemiology of the virus (Figure 3), including its geographic spread (Figure 4), was another male topic. Both relate to sharing news about the spread and extent of the virus.

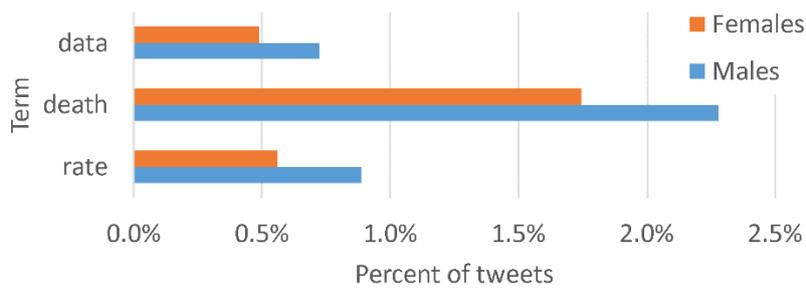

Figure 3. Epidemiology-related terms with gender differences in usage.

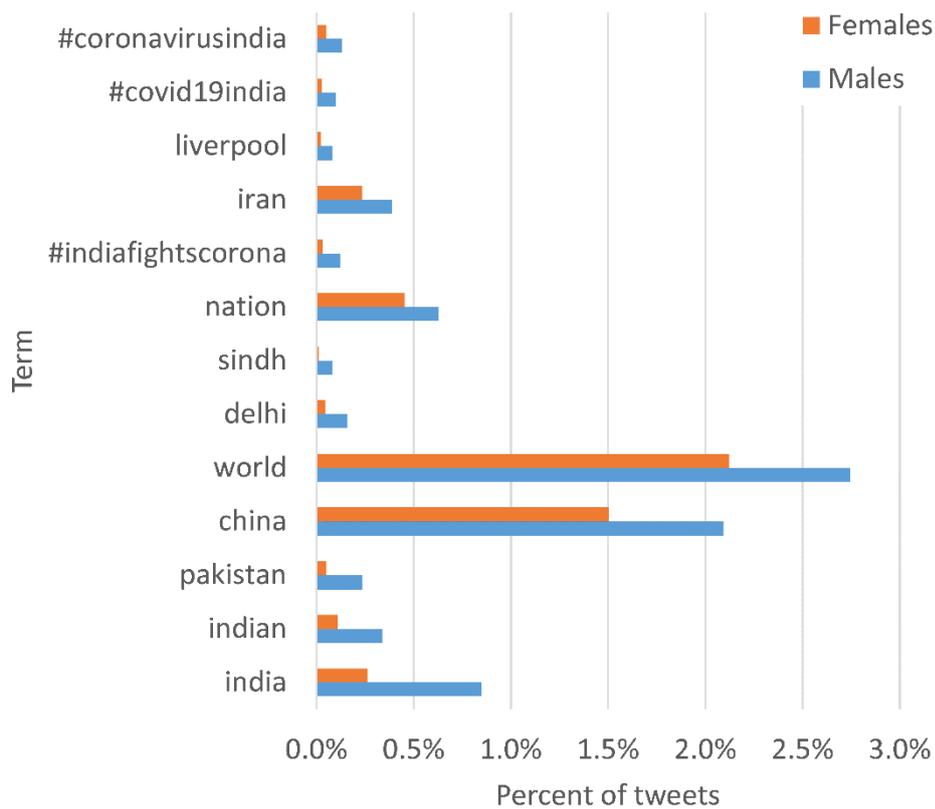

Figure 4. Geography-related terms with gender differences in usage.

## Female-oriented themes

Female-oriented themes seemed to focus on the first and second lines of defense against the virus. The key theme of social distancing is moderately female-oriented (Figure 5), in the sense that females were more likely to use the #socialdistancing hashtag and the need to stay at home as far as possible. Partly related to social distancing but also to lockdowns, females were more likely to mention family members (Figure 6) and to use all pronouns (Figure 7). Pronouns were typically used for a mix of purposes but tweets with pronouns or family members seemed more likely to discuss concrete actions or practical implications for the tweeter and the people that they know. Thus, all three themes have a practical and personal orientation. Females were also more likely to tweet about education (terms: school, student, teacher; not graphed), presumably due to its impact on themselves or their family.

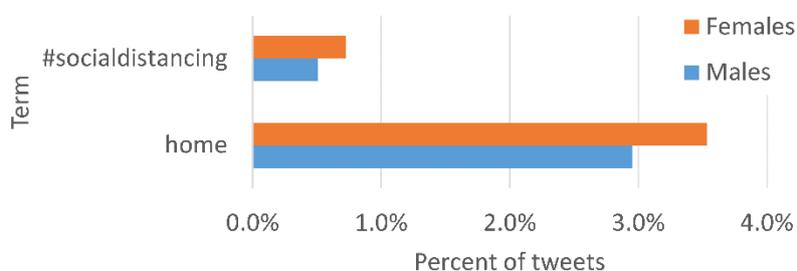

Figure 5. Social distancing-related terms with gender differences in usage.

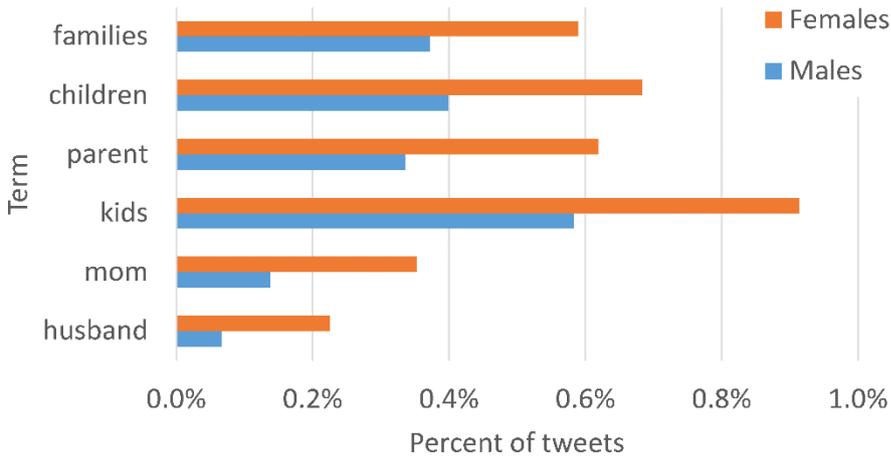
Figure 6. Family -related terms with gender differences in usage.

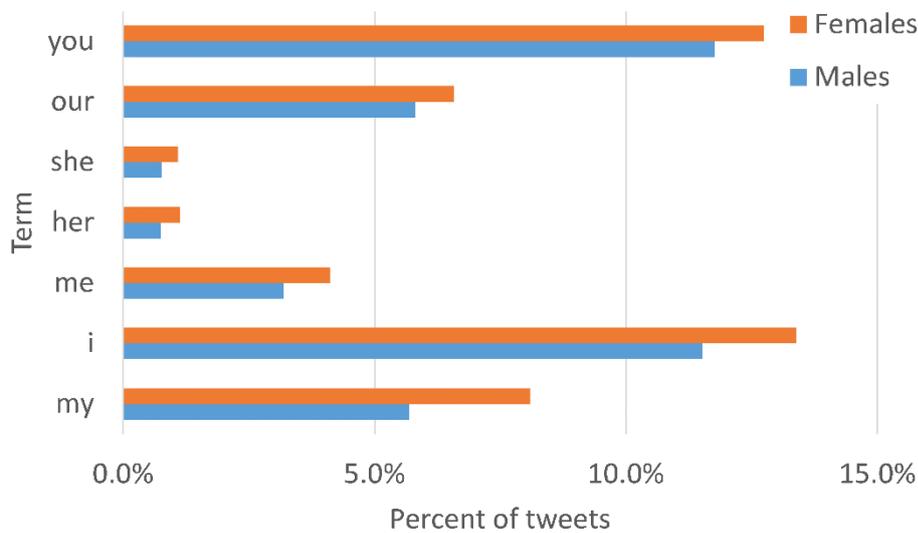
Figure 7. Pronouns with gender differences in usage.

Females were also more likely to discuss healthcare issues (Figure 8). These tweets were less focused on immediate practical issues but on the main line of defense against the virus, should the practical steps fail. Related to this, females were also more likely to express gratitude to healthcare workers and others (terms not graphed) and anxiety (see below).

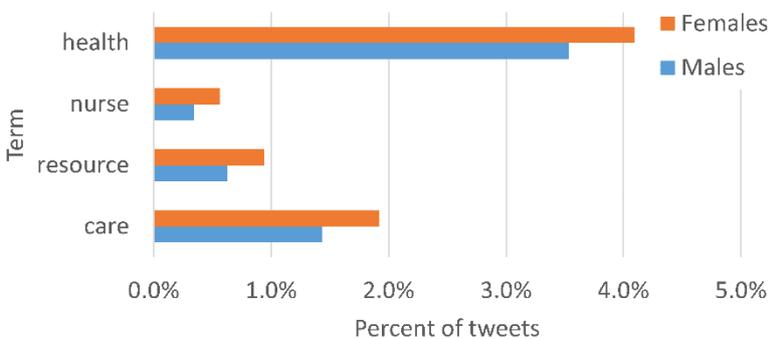
Figure 8. Healthcare-related terms with gender differences in usage.

*Mixed gender*

Two broad themes were mixed gender in the sense of some terms being male-associated and others being female-associated. Males were more likely to discuss the virus as a war whereas females were more likely to mention their anxiety about its effects (Fig 9). The war metaphor is a way of generalizing the situation as well as perhaps for males glamorizing actions against it, or emphasizing the seriousness of the issue. Thus, war metaphors could be an indirect way of expressing anxiety.

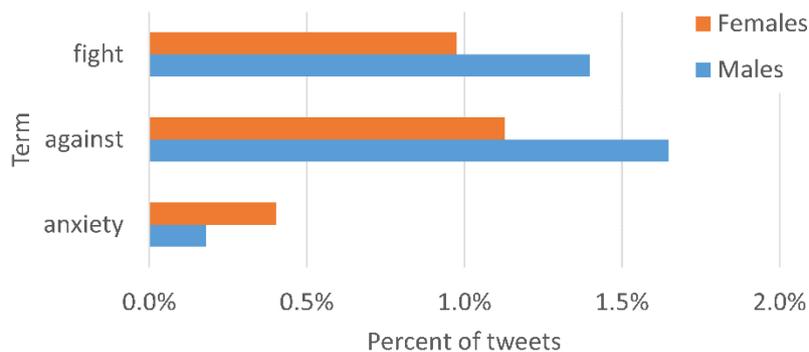

Figure 9. Fight or worry-related terms with gender differences in usage.

There were mixed gender differences in discussions of curfews (Figure 10). Whilst males were more likely to announce the existence of a curfew, females were more likely to discuss its practical impacts.

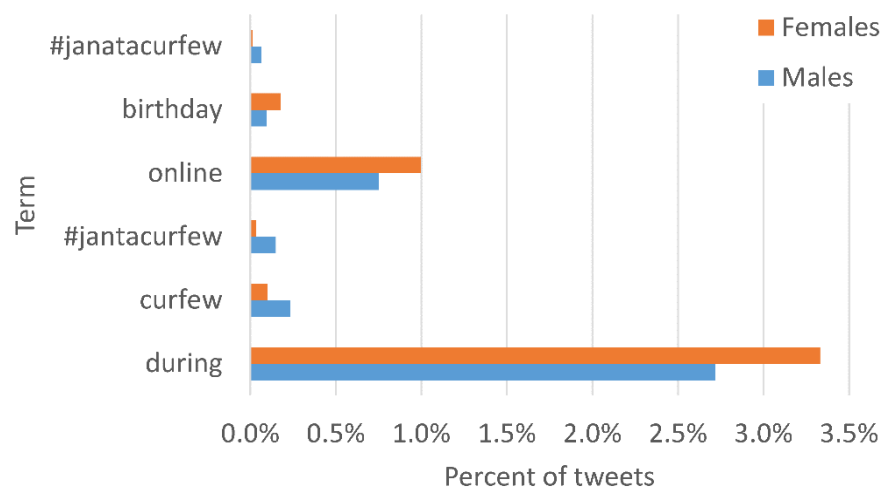

Figure 10. Curfew-related terms with gender differences in usage.

## Discussion

This quick analysis of gender differences in English tweeting about COVID-19 has several limitations. In addition to the issues discussed above, another important aspect is that Twitter does not report the geographic location of the tweets and so the data has unknown origins. In particular, if some countries have an unusually high proportion of active tweeters of one gender, then this could translate into tweets about that country statistically significantly associating with that gender with the tests used above.

The results are broadly consistent with previous research into gender differences in language use, including on social media, and gender differences in interests. The primary contribution here is therefore to so show which gender differences translate to COVID-19 on Twitter, rather than finding new gender differences.

The greater male interest in sport in many countries is widely known (e.g., Plaza, Boiché, Brunel, & Ruchaud, 2017), and males also seem to discuss politics more (or at least more directly: Bode, 2017). The greater female focus on caring (65% of family caregivers are female in the USA: Family Caregiver Alliance, 2019), and family (Parker, Horowitz, & Rohal, 2015) has also been found before. In terms of language use, females have often been found to use personal pronouns more in some types of text (Argamon, Koppel, Fine, & Shimoni, 2003).

## Conclusions

Although these conclusions are drawn from statistical tests on big data from Twitter, inferences from the results are tentative due to the processing limitations above that could not be addressed and the lack of evidence connecting offline actions to the content of tweets. Thus, for example, the greater female tendency to tweet about families does not prove that females were more concerned about the welfare of their families due to COVID-19, although this is a plausible explanation. Thus, the conclusions should be treated similarly to those of purely qualitative research: as evidence-based ideas but not proof of those ideas.

The substantially greater focus of males on sport in tweets about COVID-19 might be taken as evidence that males were less serious about the disease in the initial stages. Irrespective of whether this is true, sport was an important factor in the reaction to COVID-19 for many males. A policy-related suggestion from this is that cancelling sporting events may be particularly effective in communicating to males the seriousness of a situation. For example, if the population is told to socially distance but allowed to attend mass sporting events on the basis that an alternative (watching the event in crowded pubs or bars) is more dangerous then this may send mixed messages since crowded sporting events clearly involve close proximity with large numbers of strangers. Thus, any relaxation of bans on sporting events should be considered very carefully in the future, in countries where they are in place, and sporting bans should be considered in other countries as an important component of social distancing strategies, both for the spreading risk and the message sent to (mainly) males.

The results are consistent with, but do not prove, that women are at the forefront of actions to prevent the spread of COVID-19. Public health messages might therefore need to be particularly careful that core messages are transmitted effectively to women in media that they consume so that social distancing is fully understood by as many as possible so that it can be carried out as effectively as possible.